\documentclass[twocolumn,prb,aps,10pt]{revtex4-1}

\usepackage{amsfonts,amssymb}
\usepackage[sumlimits,intlimits]{amsmath}
\usepackage{graphics}
\usepackage{graphicx}
\usepackage{epsfig}
\usepackage{mathrsfs}
\usepackage{wasysym}
\usepackage{textcomp}
\usepackage{verbatim}
\usepackage{hyperref}    % links
\usepackage{bm}          % bold math letters

\pdfoutput=1

\newcommand{\be}{\begin{equation}}
\newcommand{\ee}{\end{equation}}
\newcommand{\Tes}{T_\text{ES}}
\newcommand{\e}{\varepsilon}
\newcommand{\sgn}{\operatorname{sgn}}

\begin{document}

\title{Coulomb gap triptychs, $\sqrt{2}$ effective charge, and hopping transport in periodic arrays of superconductor grains}

\author{Tianran Chen}
\author{Brian Skinner}
\author{B. I. Shklovskii}

\affiliation{Fine Theoretical Physics Institute, University of Minnesota, Minneapolis, Minnesota 55455}

\date{\today}

\begin{abstract}

In granular superconductors, individual grains can contain bound Cooper pairs while the system as a whole is strongly insulating.  In such cases the conductivity is determined by electron hopping between localized states in individual grains.  Here we examine a model of hopping conductivity in such an insulating granular superconductor, where disorder is assumed to be provided by random charges embedded in the insulating gaps between grains.  We use computer simulations to calculate the single-electron and electron pair density of states at different values of the superconducting gap $\Delta$, and we identify ``triptych" symmetries and scaling relations between them.  At a particular critical value of $\Delta$, one can define an effective charge $\sqrt{2}e$ that characterizes the density of states and the hopping transport.  We discuss the implications of our results for magnetoresistance and tunneling experiments.

\end{abstract}
\maketitle

\section{Introduction}

Granular superconductors are arrays of superconducting granules that are connected by electron tunneling.  As such, these systems combine the unique electronic spectrum of superconducting quantum dots with the strong Coulomb correlations that are ubiquitous in disordered systems \cite{Beloborodov2007ges}.  Among the more celebrated properties of granular superconductors are a giant magnetoresistance peak \cite{Steiner2005sii, Baturina2008hri, Lin2011mqp} and a superconductor-insulator transition that can be tuned by disorder or magnetic field \cite{Gerber1997iti, Sherman2012mos}.  So far, a comprehensive theory of the electron conductivity that can explain these features remains elusive.

In this paper, we focus on the strongly disordered limit, where the array of superconducting grains as a whole is insulating while individual grains may still retain prominent features of superconductivity.  In this case, electronic states are localized and electron conduction proceeds by phonon-assisted tunneling, or ``hopping," of electrons between grains through the insulating gaps which separate them.  In principle, electronic conduction can occur either through tunneling of single electrons or through simultaneous tunneling of an electron pair.  In this introductory discussion we concentrate primarily on hopping by single electrons; hopping by electron pairs is discussed more fully in the body of the text.  Here we note only that coherent tunneling of Cooper pairs (the Josephson effect) is neglected throughout this paper, since it is not relevant in the strongly disordered limit that we are considering.

Since hopping conductivity is a thermally activated process, its magnitude at a given temperature $T$ depends on two important energy scales associated with the spectrum of electron energy states within each grain.  The first is the charging energy $E_c = e^2/2C_0$, where $e$ is the electron charge and $C_0$ is the self-capacitance of a single grain.  The importance of the charging energy can be seen by considering that, in a neutral system, conduction requires an electron to hop from one neutral grain to another, thereby producing two charged grains, each with Coulomb self-energy $E_c$.  The second important energy scale is the superconducting gap $\Delta$, which represents an activation energy for separating a Cooper pair.  In the limit where $\Delta/E_c \rightarrow 0$, the array is equivalent to a granular metal \cite{Beloborodov2007ges, Talapin2010poc, Zhang2004dos, Chen2012cgt}.  In the opposite limit, where $\Delta/E_c \rightarrow \infty$, each grain has the properties of a bulk superconductor.  In this paper our focus is on exploring the novel physics that results when $E_c$ and $\Delta$ are similar in magnitude.

Since the superconducting gap $\Delta$ is typically on the order of $1$ meV or smaller \cite{Black1996sos, Sherman2012mos}, $E_c \sim \Delta$ implies that the self-capacitance $C_0 \gtrsim 80$ aF.  This relatively large self-capacitance can be achieved either by fabricating large grains or by surrounding the grains by an environment with a high effective dielectric constant $\kappa$, so that the product of $\kappa$ and the grain diameter $D$ satisfies $\kappa D \gtrsim 400$ nm.  For three-dimensional (3d) arrays, large $C_0$ can also be achieved by making an array of very densely-packed grains, for example, cubic grains separated by a thin insulating layer \cite{Zhang2004dos}.  In this paper we assume that the Josephson coupling energy $J$ between grains satisfies $J \ll E_c$, so that the array is indeed insulating \cite{Beloborodov2007ges} regardless of the value of $\Delta$, and coherent tunneling of Cooper pairs is absent.  Since we are considering the case of relatively large grains, we also assume that the spacing $\delta$ between discrete electron energy eigenstates within the grain satisfies $\delta \ll E_c$.

In relatively dense arrays, electron conduction can occur both through hopping of electrons between nearest-neighboring grains and through ``cotunneling" of electrons between distant grains via a chain of intermediate virtual states in intervening grains \cite{Moreira2011ect, Shklovskii1984npi, Beloborodov2007ges, Zhang2004dos}.  In the presence of some disorder, the latter mechanism dominates at low temperatures, where the length of hops grows to optimize the conductivity. This transport mechanism, called variable range hopping (VRH), was introduced by Mott \cite{Mott1968cig} and is ubiquitous in low-temperature systems with localized electron states.  

The optimum average hop length at a given temperature depends on the distribution of available ground state energies among grains in the system, called the ``density of ground states" (DOGS).  For systems with unscreened Coulomb interactions, it can be shown on the basis of a very general stability criterion of the global ground state that the DOGS must vanish \cite{Efros1975cga} at the Fermi level $\mu$.  This vanishing DOGS is called the Coulomb gap, and in the canonical Coulomb glass model of disordered systems it leads to a conductivity $\sigma$ that obeys the Efros-Shklovskii (ES) law:
\be
\sigma \propto \exp[-(\Tes/T)^{1/2}], \label{eq:ES}
\ee
where
\be 
\Tes = \frac{C e^2}{\kappa\xi}
\label{eq:Tes}
\ee
is a characteristic temperature, $C$ is a numerical coefficient, and $\xi$ is the electron localization length of the array.  Eq.\ (\ref{eq:ES}) has been observed in a number of granular metals and superconductors at low temperature (see Ref.\ \onlinecite{Beloborodov2007ges} and references therein).

In a previous paper \cite{Chen2012cgt}, we used a computer simulation to explore VRH in two-dimensional (2d) and 3d arrays of monodisperse normal metallic grains.  In such systems disorder is provided by donors and acceptors randomly situated in the interstitial spaces between grains---for example, in the metal oxide of the grains. We showed that as a consequence of the periodic charging spectrum of individual grains there is not one but three identical adjacent Coulomb gaps in the DOGS (one full gap at the Fermi level and two ``half-gaps" on either side), which together form a structure that we termed a ``Coulomb gap triptych."  Unlike in conventional Coulomb glass models, in metallic granular arrays the DOGS has a fixed width in the limit of large disorder.

\begin{figure}[tb!]
\centering
\includegraphics[width=0.48 \textwidth]{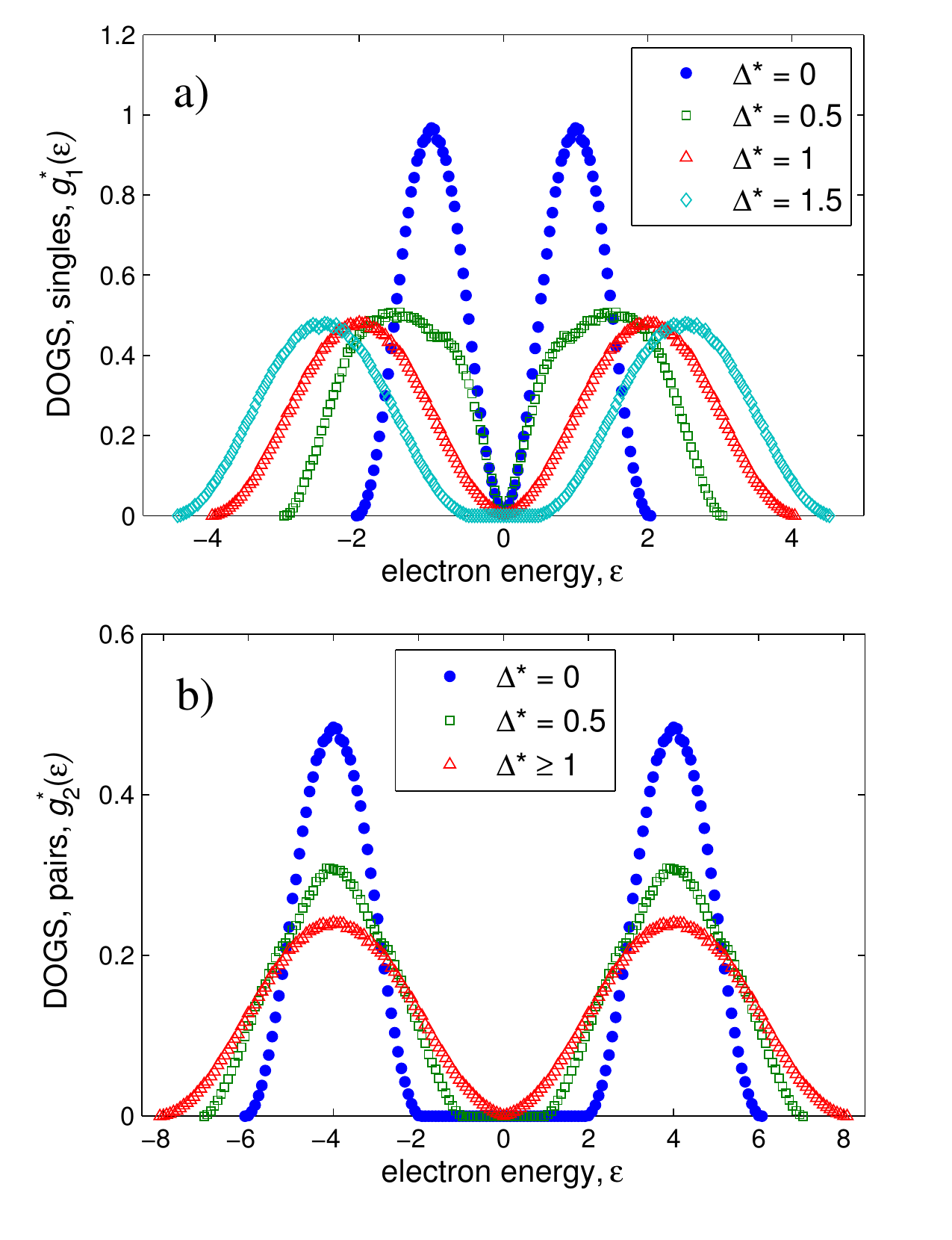}
\caption{(Color online) Single electron and pair DOGS, $g_1^*(\e)$ and $g_2^*(\e)$, of a regular 2d array of monodisperse metallic grains as a function of the dimensionless electron energy $\e = E/E_c$ at different values of the superconducting gap $\Delta^* = \Delta/E_c$. At $\Delta^* < 1$, the single electron DOGS $g_1^*$ has a soft Coulomb gap at $\e=0$, while the pair DOGS $g_2^*$ has a hard gap, and the situation is reversed for $\Delta^* > 1$. $\Delta^* = 1$ is a critical point at which both $g_{1,2}^*$ have a soft Coulomb gap. The three DOGS curves corresponding to $g^*_1(\e)$ at $\Delta^* = 0, 1$ and $g^*_2(\e)$ at $\Delta^* \geq 1$ constitute ``Coulomb gap triptychs" and can be scaled onto each other by rescaling the electron charge, as discussed in Sec.\ \ref{sec:results}.  One can equivalently say that these three curves exhibit effective charges $e$, $\sqrt{2}e$, and $2e$, respectively.}
\label{fig:DOGS}
\end{figure}

This previous study can be considered as a model for a granular superconductor in the limit where $\Delta/E_c \rightarrow 0$.  In the present paper, we generalize the theory of Ref.\ \onlinecite{Chen2012cgt} to the case of finite $\Delta^* \equiv \Delta/E_c$.  Specifically, we assume that within each grain electrons can form Cooper pairs, thereby lowering the system energy by $-2 \Delta$ per pair.  We use this model to study the DOGS and conductivity as a function of the gap $\Delta$ and the temperature $T$.  Our result for the DOGS of single electrons, which we denote $g_1(E)$, is shown in Fig. \ref{fig:DOGS}a for different values of $\Delta^*$. Fig.\ \ref{fig:DOGS}b shows the DOGS for electron pairs, $g_2(E)$.  In both cases the dimensionless energy $\e \equiv E/E_c$ is defined relative to the Fermi level, so that states with $\e < 0$ are occupied and states with $\e > 0$ are empty.  These results for the DOGS are explained more completely below, but here we note briefly that for sufficiently large disorder the system exhibits repeated Coulomb gaps at all values of $\Delta$, and the width of the DOGS is determined by the values of $E_c$ and $\Delta$ and not by the strength of the disorder.  

For energies close to the Fermi level our results for $g_1(E)$ and $g_2(E)$ are similar to those of an earlier seminal work \cite{Mitchell2012tcg}, which aimed to capture the effect of pairwise attraction of electrons on the Coulomb gap and VRH conductivity. The authors of Ref.\ \onlinecite{Mitchell2012tcg} started from the canonical Efros model of the Coulomb glass \cite{Efros1976cgi} with strong disorder and added the possibility of occupation of a site by two electrons with a finite (positive or negative) interaction energy $U$.  They used this model to study how varying the on-site energy $U$ affects $g_1(E)$, $g_2(E)$, and the hopping conductivity $\sigma$ in the presence of large external disorder.  In the present paper we examine a model that is more realistic for granular superconductors, and we confirm a number of interesting observations made in Ref. \onlinecite{Mitchell2012tcg}.  

The most interesting of these observations is the appearance of effective charges $e^* = 1e$, $e^* = 2e$, and $e^* = \sqrt{2}e$ at $\Delta^* < 1$, $\Delta^* > 1$, and $\Delta^* = 1$, respectively.  These effective charges play a prominent role in the DOGS and the hopping conductivity, as we show below.  The unusual $\sqrt{2}e$ charge that arises at $\Delta^* = 1$ is the product of a degeneracy in the electronic spectrum, and can be seen as a consequence of singly-charged electrons hopping in a Coulomb landscape whose properties are determined by doubly-charged Cooper pairs \cite{Mitchell2012tcg}.

The remainder of this paper is organized as follows.  In Sec.\ \ref{sec:model} we define the system being studied and outline our simulation technique.  In Sec.\ \ref{sec:results} we describe our main results for the DOGS and conductivity, focusing primarily on 2d arrays, and we present their implications for magnetoresistance.  In Sec.\ \ref{sec:3d} we show that our results generalize to the 3d case as well.  Sec.\ \ref{sec:tunneling} is devoted to translating our results for the DOGS into a prediction for tunneling experiments.  We close in Sec.\ \ref{sec:conclusion} with concluding remarks.

\section{Model} \label{sec:model}

In this paper we consider an array of identical, spherical grains with diameter $D$ arranged in a regular, $d$-dimensional square lattice with lattice constant $D' > D$.  For simplicity of discussion, during the majority of this paper we focus on case $d = 2$; results for $d = 3$ are presented in Sec.\ \ref{sec:3d}.  Disorder in this system is assumed to be provided by impurity charges $\pm e$ that are embedded in the insulating interstitial spaces between grains.  Such impurity charges can be thought to effectively create a random fractional donor charge $Q_i$ that resides on each grain $i$, for reasons that are explained more fully in Ref.\ \onlinecite{Chen2012cgt}. The net charge of the grain can then be written as $q_i = Q_i - en_i$, where $n_i$ is the integer number of electrons that reside on the grain relative to its neutral state.  We emphasize that $n_i$ can be a positive or negative integer, and can be defined as $n_i = N_i - I_i$, where $I_i$ is the number of positive ions and $N_i$ the number of electrons at grain $i$.  Within each grain, the $N_i$ electrons can form bound pairs through the local attraction energy $\Delta$.  In general, $\Delta$ can be tuned by an applied magnetic field $B$, as discussed below.  

Given this model, the Hamiltonian for the system can be written
\begin{eqnarray}
H & = & \sum_i \frac{(Q_i - en_i)^2 }{2 C_0} + \sum_{\langle i,j \rangle } C^{-1}_{ij} (Q_i - en_i)(Q_j - en_j) \ \ \ \ \ \\  \nonumber
 & & - 2 \Delta \sum_i  \left \lfloor \frac{N_i}{2}\right \rfloor.
\label{eq:H}
\end{eqnarray}
Here, the first term describes the Coulomb self-energy of each grain and the second term describes the Coulomb interaction between charged grains.  The coefficient $C^{-1}_{ij}$ is the inverse of the matrix of electrostatic induction $C_{ij}$. In the numerical simulations that we describe below, we make the approximations $C_0 = \kappa D/2$ and $C^{-1}_{ij} = 1/\kappa r_{ij}$.  The third term in the Hamiltonian describes the total pairing energy of electrons;  $N_i$ is the number of electrons and $\lfloor N_i/2 \rfloor$ is the number of electron pairs within grain $i$.  In Ref.\ \onlinecite{Mitchell2012tcg}, the authors proposed a similar Hamiltonian as a model for disordered superconducting films such as InO$_x$.  Unlike in Eq.\ (\ref{eq:H}), the model considered in Ref.\ \onlinecite{Mitchell2012tcg} assumes that the electron occupation numbers are restricted to $N_i = 0, 1, 2$ and that disorder is provided by random, uncorrelated site energies rather than by the random charges $Q_i$.  While we consider our model more realistic for granular superconductors, we will show that it reproduces many of the features reported in Ref.\ \onlinecite{Mitchell2012tcg}. 

Because of the presence of the impurity charges, electrons become redistributed among grains from their neutral state in order to screen the disorder Coulomb potential.  The corresponding ground state arrangement of electrons among grains plays an essential role in the conductivity, since it determines the lowest empty and highest filled electron energy levels at each grain.  In our numerical simulation described below, we search for the set $\{n_i\}$ that minimizes the Hamiltonian and use it to calculate the DOGS and the conductivity.

In conventional Coulomb glass models, the characteristic strength of the disorder is a free parameter that determines the width of the DOGS \cite{Efros1976cgi, Efros1984epo}.  One can expect that in our problem a similar role is played by the typical magnitude of the disorder charge $Q_i$, which reflects the average number of impurity charges surrounding each grain.  In fact, however, in the limit where there are many such charges one can effectively adopt a simple model in which the value of $Q_i$ is chosen randomly from the uniform distribution $Q_i \in [-e, +e]$.  To see why this model is valid, consider that each grain minimizes its Coulomb self-energy by minimizing the magnitude of its net charge, $|q_i| = |Q_i - en_i|$.  In the absence of any Cooper pairing, $n_i$ may freely take any integer value in order to arrive at a state for which $-e/2 \leq q_i \leq e/2$.  If one assumes, on the other hand, that Cooper pairing is so strong that all electrons are paired in the ground state ($\Delta^* \gg 1$) , then $N_i = n_i + I_i$ may still freely take any even-integered value, so that in the ground state $-e \leq q_i \leq e$.  In other words, regardless of the value of $\Delta$, each grain can effectively adjust to the presence of an arbitrarily strong charge disorder by changing its electron number $n_i$ so that its net charge acquires a magnitude smaller than $e$.  For this reason, in the limit of large disorder the DOGS has a fixed width, as first explained in Ref.\ \onlinecite{Chen2012cgt}.  For the results presented below, we take $Q_i$ to be randomly chosen from the uniform distribution $Q_i \in [-e, e]$.  The ion number $I_i$ is assumed to be very large, so that electrons are never completely depleted from any given grain.  $I_i$ is also taken to be even or odd with equal probability; the relevance of this choice is explained below.

In our analysis below it is convenient to introduce the following dimensionless units, which reduce the number of free variables in the problem.  We introduce the dimensionless distance between the centers of grains $i$ and $j$,
\be 
r_{ij}^* = \frac{r_{ij}}{D},
\label{eq:rstar}
\ee 
the dimensionless charge
\be 
q_i^* = \frac{Q_i - en_i}{e},
\label{eq:qstar}
\ee 
the dimensionless electron energy
\be 
\e = E/E_c,
\label{eq:estar}
\ee 
the dimensionless DOGS for single electrons and electron pairs
\be  
g_{1,2}^*(\e) = E_c D^d g_{1,2}(\e),
\label{eq:dos}
\ee
the dimensionless temperature
\be 
T^* = \frac{2 D k_B T}{E_c \xi},
\label{eq:Tstar}
\ee
and the dimensionless resistivity
\be 
\ln \rho^* = \frac{\xi}{2 D} \ln (\rho/\rho_0),
\label{eq:rhostar}
\ee 
where $\rho_0$ is a prefactor for the resistivity with a weak, power-law dependence on temperature.
We also assume that the gap between neighboring grains $D' - D \ll D$, so that $D' \simeq D$.
The problem then loses any explicit dependence on the diameter or the localization length. With these definitions, one can write the Hamiltonian of Eq.\ (\ref{eq:H}) in dimensionless units as 
\be
H^* = \sum_i q^{*2}_i + \sum_{\langle i,j \rangle } \frac{q^*_i q^*_j}{r^*_{ij}} - 2\Delta^* \sum_i  \left \lfloor \frac{N_i}{2}\right \rfloor.
\label{eq:Hstar}
\ee

If one is given the ground state electron occupation numbers $\{n_i\}$, then one can determine the highest occupied electron energy state, $\e_i^{1-}$, and the lowest empty state, $\e_i^{1+}$, at a given grain $i$.  These energies determine the contribution of the grain $i$ to the single-electron conductivity, and are given by:
\be
\e_i^{1-} = - 2q_i^* -1 - \sum_{j \neq i} \frac{q^*_j}{r^*_{ij}} - 
\left\{ \begin{array}{ll}
0 & \text{,  $N_i$ odd} \\
2 \Delta^* & \text{,  $N_i$ even}
\end{array}
\right. ,
\label{eq:sf}
\ee 
\be 
\e_i^{1+} = - 2q_i^* + 1 - \sum_{j \neq i} \frac{q^*_j}{r^*_{ij}} -  
\left\{ \begin{array}{ll}
2 \Delta^* & \text{,  $N_i$ odd} \\
0 & \text{,  $N_i$ even}
\end{array}
\right. .
\label{eq:se}
\ee

\begin{figure}[htb!]
\centering
\includegraphics[width=0.5 \textwidth]{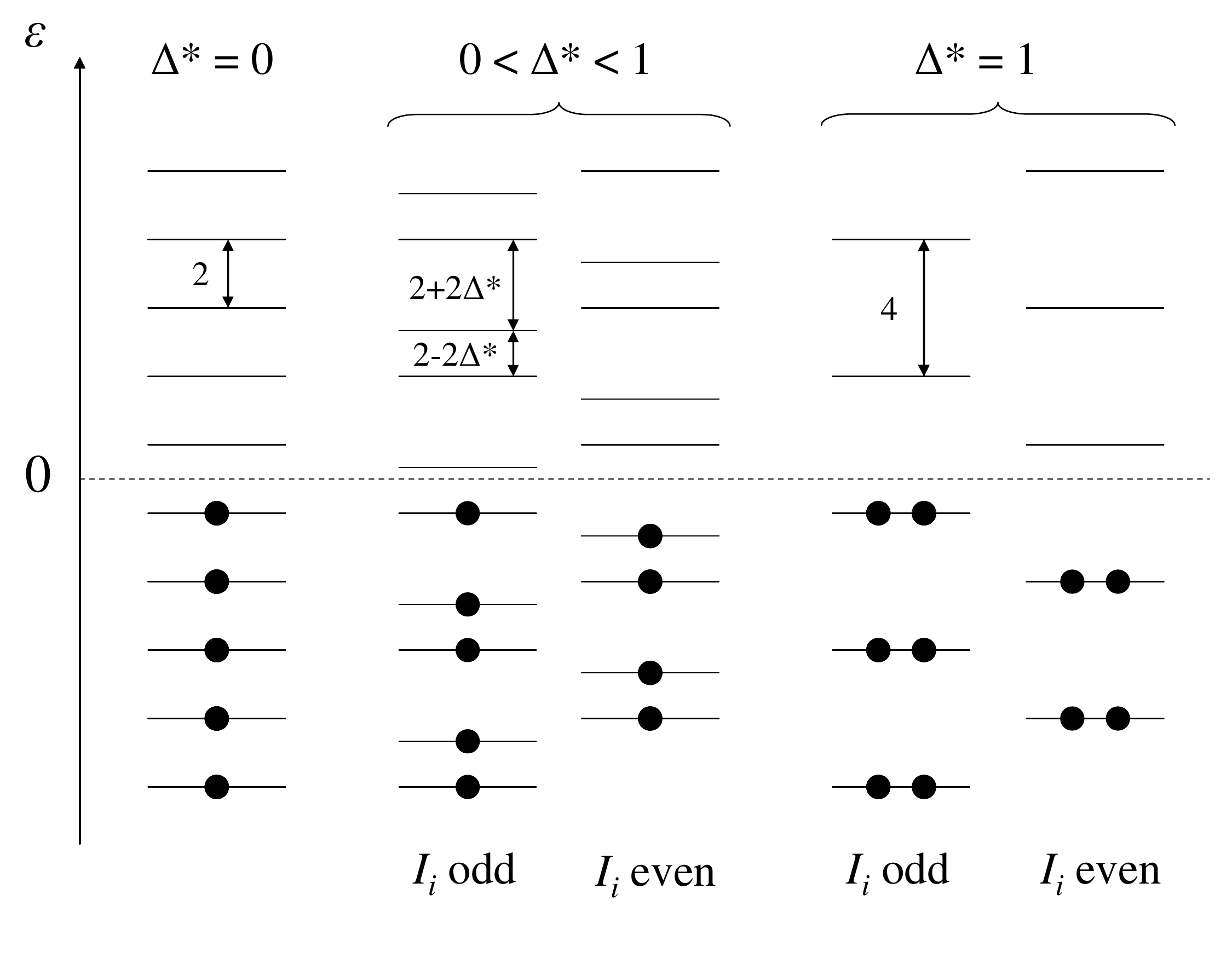}
\caption{Single-electron energy levels of an isolated neutral grain. At $\Delta^* = 0$ (left), the Coulomb self-energy produces a spectrum where different charge states $\e$ are separated in energy by $2$.  At  $0 < \Delta^* < 1$ (center), those energy levels corresponding to the addition of an electron to a grain with an odd total number of electrons are shifted by $-2 \Delta^*$.  At $\Delta^* = 1$ (right), the difference in self-energy $2 E_c$ between two successive charge states is compensated by the pairing energy $-2 \Delta$, so that pairs of subsequent electron levels merge. The Fermi level at $\e = 0$ is indicated schematically by a dashed line.}
\label{fig:spec}
\end{figure}

From Eqs.\ (\ref{eq:sf}) and (\ref{eq:se}) one can see that the spectrum of single-electron energy levels at a given grain $i$ depends on the ``parity" of the grain: whether the number of electrons in the neutral state, $I_i$, is odd or even.  To understand why this is the case, consider first the spectrum of a single grain with $\Delta = 0$.  In such a grain, $\e_i^{1+} - \e_i^{1-} = 2$ regardless of the number of electrons in the grain.  This implies a ladder of electron energy levels, spaced by $2 E_c$, corresponding to different charge states of the grain.  These energy levels are shown schematically in the left side of Fig.\ \ref{fig:spec}.  When $\Delta^*$ is finite, on the other hand, those energy states corresponding to an even total number of electrons in the grain become shifted by $-2 \Delta^*$ as a consequence of the attractive interaction between electron pairs.  As a result, $\e_i^{1+} - \e_i^{1-} = 2 - 2\Delta^*$ for grains with odd $N_i$ and $\e_i^{1+} - \e_i^{1-} = 2 + 2\Delta^*$ for grains with even $N_i$.  This suggests that the energy to add or remove one electron from the grain's neutral state depends on the parity of the grain, as shown in the center of Fig.\ \ref{fig:spec}.  (The importance of the grain parity for its electronic spectrum has been well established by previous theoretical \cite{Averin1992sco, Matveev1997pei} and experimental \cite{Tuominen1992eep, Tuominen1993een} studies.)  At $\Delta^* = 1$, pairs of electron energy states become two-fold degenerate, as shown on the right side of Fig.\ \ref{fig:spec}.  As a consequence, at $\Delta^* \geq 1$ in the ground state all grains have an even total number of electrons, regardless of the disorder strength.  This uniform pairing has an important consequence for the DOGS, as discussed below.

The diagram of Fig.\ \ref{fig:spec} shows that the spacing between energy levels at $\Delta^* = 1$ is doubled relative to that of $\Delta^* = 0$.  One can observe that this same increased spacing could be achieved if the electron charge $e$ were replaced with an effective charge $\sqrt{2}e$, so that the charging energy $E_c \propto e^2$ is doubled.  In fact, this effective charge $\sqrt{2}e$ plays a prominent role for the DOGS and conductivity at $\Delta^* = 1$, as will be shown in Sec.\ \ref{sec:results}.

In the presence of some disorder, the ladder of energy states depicted in Fig.\ \ref{fig:spec} becomes shifted randomly up or down from one grain to the next by the disorder potential.  The values of $\e_i^{1-}$ and $\e_i^{1+}$ for each grain---those energy states just above and just below the constant global Fermi level---contribute to the DOGS $g_1^*(\e)$.  

Thus far we have focused our discussion on hopping by single electrons, which is characterized by a localization length $\xi = \xi_1$.  In principle, conduction may occur through simultaneous hopping of an electron pair as well, with a distinct localization length $\xi = \xi_2$.  For example, one can expect pair tunneling to become dominant in the limit $\Delta^* \rightarrow \infty$, where thermally activated breaking of bound Cooper pairs is completely suppressed.
In order to discuss conduction by electron pairs, one can similarly define the energy associated with pair excitations, in analogy with Eqs.\ (\ref{eq:se}) and (\ref{eq:sf}).   Specifically:
\begin{eqnarray}
\e_i^{2-} = - 4q_i^* - 4 - 2 \sum_{j \neq i} \frac{q^*_j}{r^*_{ij}} - 2 \Delta^*, 
\label{eq:pf} \\
\e_i^{2+} = - 4q_i^* + 4 - 2 \sum_{j \neq i} \frac{q^*_j}{r^*_{ij}} - 2 \Delta^*. 
\label{eq:pe}
\end{eqnarray}
Note that, unlike for single electron excitations, for pairs we have $\e_i^{2+} - \e_i^{2-} = 8$ regardless of the parity of the grain.  This suggests that the ladder of energy states corresponding to pair excitations has a uniform spacing $8 E_c$, and thus all pair excitation energies are independent of $\Delta$.  This is as expected, since the total number of bound pairs in the system is unchanged by the simultaneous tunneling of a pair.  As with the single electron energy levels, the disorder potential produces a random shifting of the two-electron energy levels from one grain to another.  The energies $\e_i^{2-}$ and $\e_i^{2+}$ in the ground state are histogrammed to produce the pair DOGS, $g_2^*(\e)$.

In order to evaluate numerically the DOGS, we use a computer simulation to search for the ground state arrangement of electrons, $\{n_i\}$, in a finite array of grains.  This is done by looping over all pairs $ij$ and attempting to move either one or two electrons from $i$ to $j$. If the proposed move lowers the total system energy $H^*$, then it is accepted, otherwise it is rejected.  This process is continued until no single-electron or pair transfers are possible that lower $H^*$.  Equivalently, one can say that for all $i$, $j$ we check that two sets of ES ground state criteria are satisfied:
\be
\e_j^{1+} - \e_i^{1-} - 1/r^*_{ij} > 0.
\label{eq:ES1} 
\ee
and
\be
\e_j^{2+} - \e_i^{2-} - 4/r^*_{ij} > 0.
\label{eq:ES2}
\ee
The final arrangement of electrons can be called a ``pseudo-ground state," which is not strictly equal to the true ground state of the system but which generally provides an identical DOGS up to very small energies \cite{Efros1984epo, Mobius1992cgi, Efros2011cgi}.

Once the energies $\{\e_i^{1 \pm, 2 \pm}\}$ are known, we evaluate the resistivity using the approach of the Miller-Abrahams resistor network \cite{Miller1960ica}.  This approach is described in detail in Refs.\ \onlinecite{Chen2012cgt, Skinner2012toh}, but here we give a brief conceptual overview.  In the Miller-Abrahams description, each pair of grains $ij$ is said to be connected by some equivalent resistance $R_{ij}$.  The value of $R_{ij}$ increases exponentially with the distance $r_{ij}$ between the grains and with the activation energy $E_{ij}$ required for hopping between them according to $R_{ij} \propto \exp[2 r_{ij}/\xi + E_{ij}/k_BT]$.  Note that, using the dimensionless units of Eqs.\ (\ref{eq:rstar}) -- (\ref{eq:rhostar}), one can define the dimensionless logarithm of the resistance $\ln R_{ij}^* = r^*_{ij} + \e_{ij}/T^*$, which has no explicit dependence on $\xi$.  The resistivity of the system as a whole is found using a percolation approach.  Specifically, we find the minimum value $R_c$ such that if all resistances $R_{ij}$ with $R_{ij} < R_c$ are left intact while others are eliminated (replaced with $R = \infty$), then there exists a percolation pathway connecting opposite faces of the simulation volume.  The system resistivity is equated with $R_c D^{d-2}$.

In principle, single-electron hopping and pair hopping provide parallel mechanisms for charge transport between a given pair of grains $ij$, and so they can be represented as parallel resistors connecting the two grains.  In most situations, however, one of the two mechanisms dominates the conductivity while the other can be neglected, as we show below.  
We therefore focus primarily on the case where single and pair excitations can be treated as independent, non-connected resistor networks with resistivities $\rho_1$ and $\rho_2$, respectively.  Some limited results for mixed conduction are provided at the end of the following section.

All numerical results for $2$d systems presented in the following section correspond to simulations of $100 \times 100$ lattice sites with open boundary conditions, averaged over $1000$ independent, random realizations of the disorder.  Energies are defined relative to the Fermi level $\mu$, so that in the ground state $\e_i^{1-} < 0$, $\e_i^{2-} < 0$ and $\e_i^{1+} > 0$, $\e_i^{2+} > 0$ for all $i$.

\section{Results} \label{sec:results}

The DOGS is shown in Fig.\ \ref{fig:DOGS} for single electron excitations, $g^*_1(\e)$, and for pair excitations, $g^*_2(\e)$, at different values of the gap $\Delta^*$.  For each curve, the DOGS vanishes at the Fermi level ($\e = 0$), as required by the stability criteria of Eqs.\ (\ref{eq:ES1}) and (\ref{eq:ES2}).  One can also note that the DOGS generally becomes wider with increasing $\Delta^*$ as a consequence of the widening gaps between odd and even electron energy levels (see Fig.\ \ref{fig:spec}).  
The evolution of the DOGS with $\Delta^*$ can be understood more completely as follows.  

At $\Delta^* = 0$, the array is equivalent to a granular normal metal.  As a consequence, the curve $g_1^*(\e)$ at $\Delta^* = 0$ is identical to the one reported in Ref.\ \onlinecite{Chen2012cgt}.  The most salient feature of this curve is its ``triptych" symmetry, with two identical peaks that are symmetric about their centers.  As explained in Ref.\ \onlinecite{Chen2012cgt}, this symmetry is a result of the ES stability criterion of Eq.\ (\ref{eq:ES1}) in conjunction with the uniform spacing between electron energy levels (at $\Delta^* = 0$, $\e_i^{1+} - \e_i^{1-} = 2$ for all $i$, as shown in Fig.\ \ref{fig:spec}).  Thus, the ``soft" Coulomb gap at $\e = 0$ gets repeated identically at $\e = \pm 2$.  On the other hand, the pair DOGS, $g_2^*(\e)$, at $\Delta^* = 0$ has a hard gap at the Fermi level with width $4$.  This hard gap can be understood by considering that at $\Delta^* = 0$, Eqs.\ (\ref{eq:sf}) -- (\ref{eq:pe}) imply that $\e_i^{2 \pm} = 2 \e_i^{1 \pm} \pm 2$.  Since $\e_i^{1-} < 0$ and $\e_i^{1+} > 0$ for all $i$, we have $|\e_i^{2\pm}| > 2$, and therefore there must be a hard gap of width $4$.  Physically, one can say that the gap arises in $g^*_2(\e)$ because the charging energy $4E_c$ associated with adding two electrons to a given grain is larger in magnitude than the random Coulomb potential, which is screened effectively by the rearrangement of single electrons.  As a consequence of the relation between $\e_i^{2 \pm}$ and $\e_i^{1 \pm}$, at $\Delta^* = 0$ the two DOGS can be mapped onto each other via the relation $g_1(\e) = 2 g_2[2 \e + 2 \sgn(\e)]$.  A slightly different version of this relation was reported in Ref.\ \onlinecite{Mitchell2012tcg}.

When the pairing interaction is finite but small, $0 < \Delta^* < 1$, $g_1^*(\e)$ is unchanged very close to the Fermi level, but away from the Fermi level it becomes somewhat broadened due to the widening energy gaps between even and odd parity electron states (see Fig.\ \ref{fig:spec}).  The pair DOGS, meanwhile, retains a hard gap near the Fermi level, but the width of this gap shrinks to $4(1 - \Delta^*)$.  

In the opposite case, where the pairing interaction is strong enough that $\Delta^* > 1$, the situation is reversed.  That is, the single-electron DOGS $g_1^*(\e)$ acquires a hard gap at the Fermi level while $g_2^*(\e)$ has only a soft Coulomb gap.  This result can be understood by first noting that at $\Delta^* \geq 1$, all electrons are paired in the ground state.  This is true because at $\Delta^* > 1$ any grain with an odd number of electrons can lower its energy by acquiring an electron from a distant grain with electron energy close to the Fermi level (or from the voltage source).  Making use of Eqs.\ (\ref{eq:sf}) -- (\ref{eq:pe}) for even-parity grains produces the relation $\e_i^{1 \pm} = \frac12 \e_i^{2 \pm} \pm (\Delta^* - 1)$.  Since $\e_i^{2+} > 0$ and $\e_i^{2-} < 0$, it follows that all $|\e_i^{1\pm}| > \Delta^* - 1$ for all $i$, and thus there is a hard gap in $g_1^*(\e)$ of width $2(\Delta^* - 1)$.  Physically, this hard gap arises because the pairing interaction is stronger than the disorder Coulomb potential, which is screened effectively by Cooper pairs.  Thus, any excitation of single-electron hops requires a finite activation energy of at least $\Delta^* - 1$.  The relations between $\e_i^{2 \pm}$ and $\e_i^{1 \pm}$ at $\Delta^* > 1$ imply a mapping between $g_1^*(\e)$ and $g_2^*(\e)$ that was also noticed by Ref.\ \onlinecite{Mitchell2012tcg}, namely $g_2^*(\e) = \frac12 g_1^*[\frac12 \e + \sgn(\e) (\Delta^* - 1)]$.  At such large values of $\Delta^*$, the fixed relation $\e_i^{2+} - \e_i^{2-} = 8$ implies that $g_2^*(\e)$ becomes saturated and has a fixed width for all $\Delta^* \geq 1$.

At the point where $\Delta^* = 1$ precisely, some remarkable features emerge in the DOGS.  This might be expected by noticing the special role played by $\Delta^* = 1$ in the single-electron energy spectrum; this is the point where pairs of energy levels become degenerate (see Fig.\ \ref{fig:spec}, right).  At $\Delta^* = 1$ neither $g_1^*(\e)$ nor $g_2^*(\e)$ has a hard gap, and in fact the two DOGS can be mapped onto each other via the simple relation $g_2^*(\e) = \frac12 g_1^*(\frac12 \e)$.  In addition, there is a simple scaling relation between $g_1^*(\e)$ at $\Delta^* = 1$ and $g_1^*(\e)$ at $\Delta^* = 0$.  Namely, 
\be 
g_1^*(\e) \Bigm|_{\Delta^* = 1} = \frac{1}{2} g_1^*\left( \e/2 \right) \Bigm|_{\Delta^* = 0} = 2 g_2^*(2 \e) \Bigm|_{\Delta^* \geq 1}.  
\label{eq:sqrt2}
\ee

The second equality in Eq.\ (\ref{eq:sqrt2}) can be understood in a straightforward way.  Indeed, the second equality suggests that one can arrive at the pair DOGS at large $\Delta^*$ by taking the single-electron DOGS at $\Delta^* = 0$ and rescaling the value of the electron charge by a factor of $2$.  Replacing $e$ by an effective charge $e^* = 2e$ in the unit of energy $E_c$ produces a factor $4$ expansion of the x-axis and a factor $4$ contraction of the y-axis, which is equivalent to the second equality in Eq.\ (\ref{eq:sqrt2}).  This scaling can be expected, since for large pairing interaction $\Delta^* > 1$, all electrons are paired, and one can naturally think that only charge $2e$ objects exist in the problem.  Thus, at such large $\Delta^*$ the problem of the arrangement of electron pairs in the disorder potential is equivalent to the problem of the arrangement of single electrons in a disorder potential, with rescaled units.

The first equality in Eq.\ (\ref{eq:sqrt2}), on the other hand, is unexpected, since it implies that $g_1^*(\e) \bigm|_{\Delta^* = 1}$ can be determined from $g_1^*(\e) \bigm|_{\Delta^* = 0}$ by replacing the electron charge with an effective charge $e^* = \sqrt{2} e$.  This remarkable feature of the DOGS at $\Delta^* = 1$ was first pointed out by Ref.\ \onlinecite{Mitchell2012tcg}.  Those authors showed that the result $e^* = \sqrt{2}e$ is the natural consequence of single electrons hopping in a Coulomb landscape that is shaped predominantly by Cooper pairs.  More formally, one can say that the pair stability criterion of Eq.\ (\ref{eq:ES2}) produces a stronger constraint on $g_1^*(\e)$ than the single-electron criterion of Eq.\ (\ref{eq:ES1}).  This can be seen by substituting $\e_i^{2 \pm} = 2\e_i^{1\pm}$, which is correct at $\Delta^* = 1$ (see Fig.\ \ref{fig:spec}), into Eq.\ (\ref{eq:ES2}).  As a result, one finds that $\e_i^{1+} - \e_j^{1-} - 2/r^*_{ij} > 0$, or in dimensionfull units, $E_i^{1+} - E_j^{1-} - (\sqrt{2}e)^2/\kappa r_{ij} > 0$.  Repeating the traditional derivation of the Coulomb gap $\cite{Efros1975cga}$ starting with this inequality leads to an effective charge $e^* = \sqrt{2}e$ in the DOGS.

In addition to its importance for the DOGS, the effective charge $e^*$ plays a prominent role in the electron conductivity.  Specifically, it enters the characteristic temperature $\Tes$ in the ES law [Eq.\ (\ref{eq:ES})].  Since $\Tes \propto e^2$ [see Eq.\ (\ref{eq:Tes})], the arguments above suggest that if one defines the ES temperature $\Tes^s(\Delta^*)$ for single-electron conductivity and $\Tes^p(\Delta^*)$ for pair conductivity at a given value of $\Delta^*$, then these should satisfy
\be 
\Tes^p(\Delta^* \geq 1) = 2 \Tes^s(\Delta^* = 1) = 4 \Tes^s(\Delta^* = 0).
\label{eq:Tesratio}
\ee

\begin{figure}[htb!]
\centering
\includegraphics[width=0.48 \textwidth]{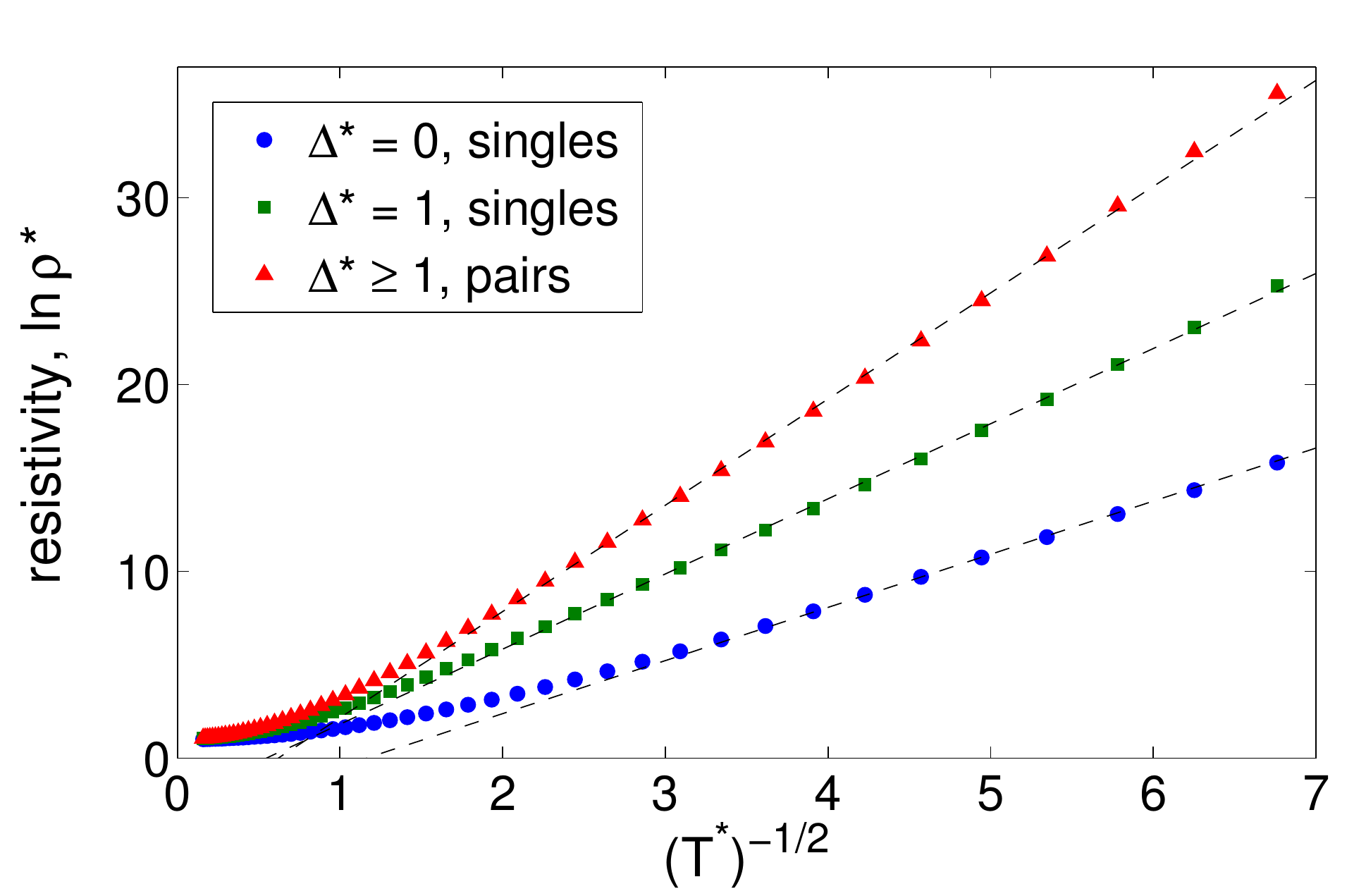}
\caption{(Color online) The temperature dependence of the resistivity for single electron conduction at $\Delta^* = 0, 1$ and pair conduction at $\Delta^* \geq 1$.  The dimensionless resistance $\ln \rho^*$ is plotted against $(T^*)^{-1/2}$ to illustrate that the resistivity follows the ES law [Eq.\ (\ref{eq:ES})] at low temperatures.  The dashed lines are linear fits whose slopes have the ratio $1$:$\sqrt{2}$:$2$.}
\label{fig:Rratio}
\end{figure}

In order to verify this prediction, we measured the single-electron resistivity $\ln \rho^*_1$ and the electron pair resistivity $\ln \rho^*_2$ at various values of $\Delta^*$ and over a range of temperatures using the resistor network approach described in Sec.\ \ref{sec:model}.  The result is plotted in Fig.\ \ref{fig:Rratio} as $\ln \rho^*$ versus $(T^*)^{-1/2}$.  As expected, at low temperature, $T^* \ll 1$, the conductivity is well described by the ES law in all cases.  By making linear best fits to the data at low temperature, we find that the corresponding temperatures $\Tes$ indeed satisfy Eq.\ (\ref{eq:Tesratio}).  This can be seen from the dashed lines in Fig.\ \ref{fig:Rratio}, which show three fit lines with relative slopes $1:\sqrt{2}:2$, as predicted by the corresponding effective charges $e^*$.  If the data in Fig.\ \ref{fig:Rratio} are fitted with independent best fit lines, we find that $\Tes^p(\Delta^* \geq 1) \approx 2.2 \Tes^s(\Delta^* = 1) \approx 4.3 \Tes^s(\Delta^* = 0) $, which is within our numerical uncertainty of the prediction in Eq.\ (\ref{eq:Tesratio}).

The evolution of $\Tes$ with $\Delta$ suggests an interesting mechanism for the magnetoresistance of the sample.  Generally speaking, the pairing energy $\Delta$ in a superconducting material decreases monotonically \cite{Tinkham1996its} with the intensity of an applied magnetic field $B$.  Thus, by applying a magnetic field one can tune the pairing energy and thereby alter the DOGS, the ES temperature $\Tes$, and the resistivity.  In the following discussion we assume that this tuning of $\Delta$ is the primary role of an applied magnetic field, and we ignore the effect of the magnetic field on hopping interference phenomena \cite{Shklovskii1991sai}.  One could also imagine that the magnetic field is applied parallel to the array, so that all hopping trajectories encircle zero magnetic flux.

In order to investigate this mechanism for magnetoresistance, we consider first the case where all conduction is due to single electron hopping.  This would be the case, for example, when $\xi_2/\xi_1 \ll 1$.  In such a case the results of Figs.\ \ref{fig:DOGS} and \ref{fig:Rratio} imply a monotonic negative magnetoresistance.  That is, as a magnetic field $B$ is applied, the gap $\Delta$ decreases, leading to a larger DOGS near the Fermi level and thus to enhanced conductivity.  More specifically, if the superconducting gap is large enough that at zero magnetic field $\Delta^* > 1$, then in the absence of a magnetic field the single-electron DOGS has a hard gap.  This hard gap implies that at low temperatures $T^* \ll (\Delta^* - 1)$, the resistivity is very large and described by an Arrhenius-type activation law.  When $B$ is increased to the point that $\Delta^* = 1$, the resistivity becomes smaller and obeys the ES law with a characteristic temperature $\Tes^s(1)$.  As the magnetic field is increased even further, $\Tes$ decreases and the resistivity declines.  This decline continues until such large fields that $\Delta^* \ll 1$, when the resistivity plateaus and $\Tes = \Tes^s(0)$.  According to the second equality in Eq.\ (\ref{eq:Tesratio}), at small temperatures one should expect that the large-$B$ resistivity and the resistivity at $\Delta^* = 1$ are related by $[\ln \rho^*_1(\Delta^* = 1)]/[\ln \rho^*_1(\Delta^* = 0)] \simeq \sqrt{2}$.

\begin{figure}[htb!]
\centering
\includegraphics[width=0.45 \textwidth]{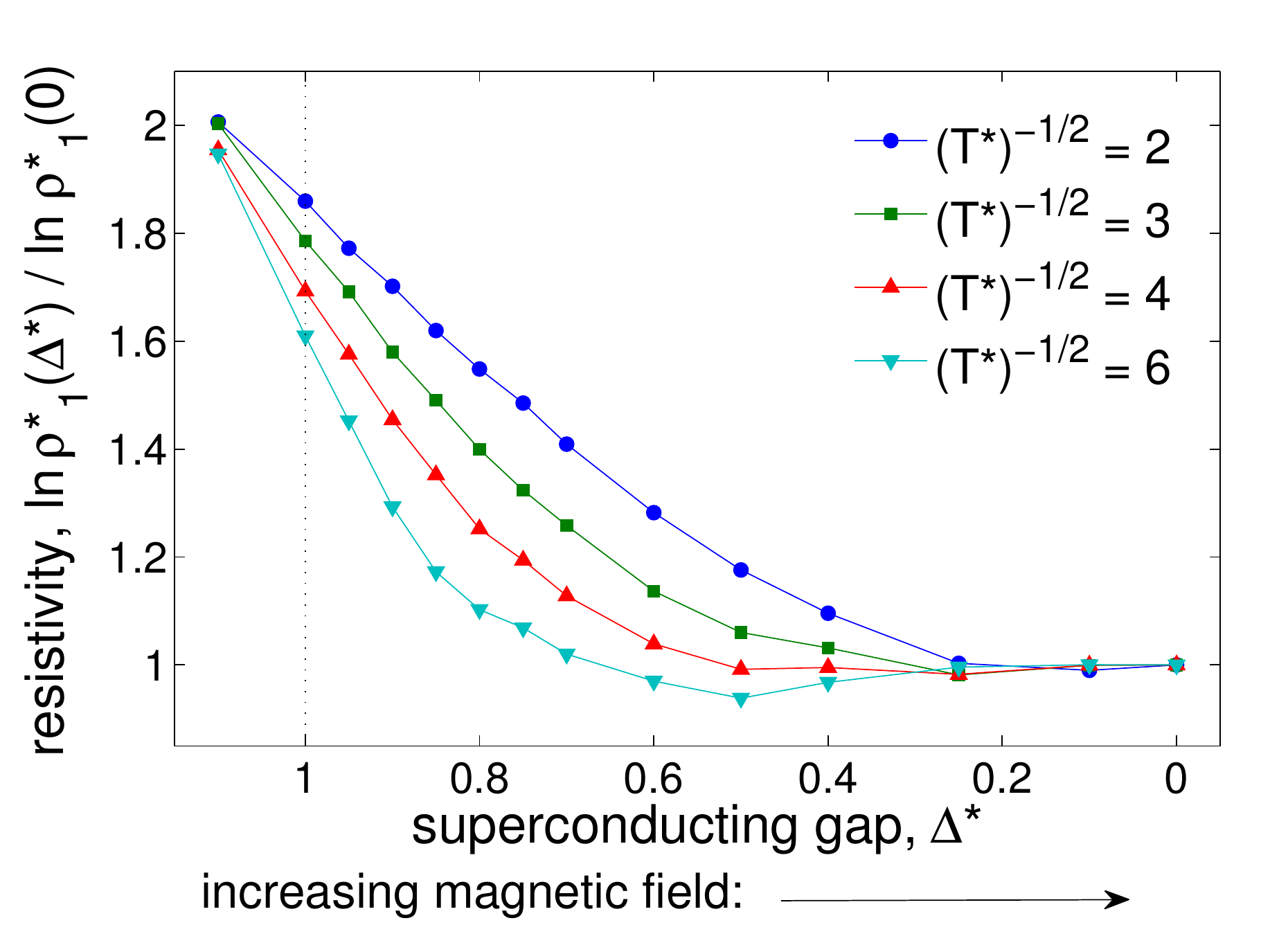}
\caption{(Color online) Resistivity for single-electron hopping, $\ln\rho_1^*$, as a function of the superconducting gap $\Delta^*$ at different values of the temperature $T^*$.  The resistivity is normalized by its value at $\Delta^*=0$, and one can see that for small temperatures the ratio $[\ln \rho^*_1(\Delta^* = 1)]/[\ln \rho^*_1(\Delta^* = 0)]$ seems to approach $\sqrt{2}$.  The declining resistivity with decreasing gap implies a negative magnetoresistance.  The dotted vertical line indicates $\Delta^* = 1$, which can be thought of as the point where the resistivity crosses over from an activated dependence to the ES law with increasing magnetic field (at small temperature).}
\label{fig:singlesR}
\end{figure}

This result is confirmed in Fig.\ \ref{fig:singlesR}, which shows the single-electron resistivity as a function of the superconducting gap $\Delta^*$ at various values of temperature.  As expected, the resistivity indeed declines with decreasing gap (increasing $B$), and at very small temperatures (large $(T^*)^{-1/2}$) the relation $[\ln \rho^*_1(\Delta^* = 1)]/[\ln \rho^*_1(\Delta^* = 0)] = \sqrt{2}$ is nearly satisfied.  This result provides an additional confirmation of the picture of an effective electron charge $\sqrt{2}e$ at $\Delta^* = 1$.

We would like to emphasize that this mechanism for negative magnetoresistance is quite unusual, and cannot be understood simply as a reduction of some activation energy due to weaker Cooper pairing.  Rather, the negative magnetoresistance arises because decreased $\Delta^*$ leads to a DOGS $g_1^*(\e)$ that is less depleted by intimidation by Cooper pairs, and thus to enhanced electron conduction at low temperature.

The results of Fig.\ \ref{fig:singlesR} focus on the case where conduction is provided by single electrons only, which is appropriate when $\xi_2/\xi_1 \ll 1$.  On the other hand, when the localization lengths $\xi_1$ and $\xi_2$ are similar in magnitude, the conduction should be dominated by single-electron hopping at $\Delta^* \ll 1$ and by pair hopping at $\Delta^* \gg 1$.  This is the case because at all $\Delta^* \neq 1$ one of the two DOGS has a hard gap.
By increasing the magnetic field, then, one can apparently produce a transition between pair-dominated conduction and single-electron-dominated conduction, provided that $\Delta^* > 1$ in the absence of applied field.  Such a transition may help to explain the giant magnetoresistance peak seen in experiments \cite{Steiner2005sii, Baturina2008hri, Lin2011mqp}, as was proposed by Ref.\ \onlinecite{Mitchell2012tcg}.

\begin{figure}[htb!]
\centering
\includegraphics[width=0.45 \textwidth]{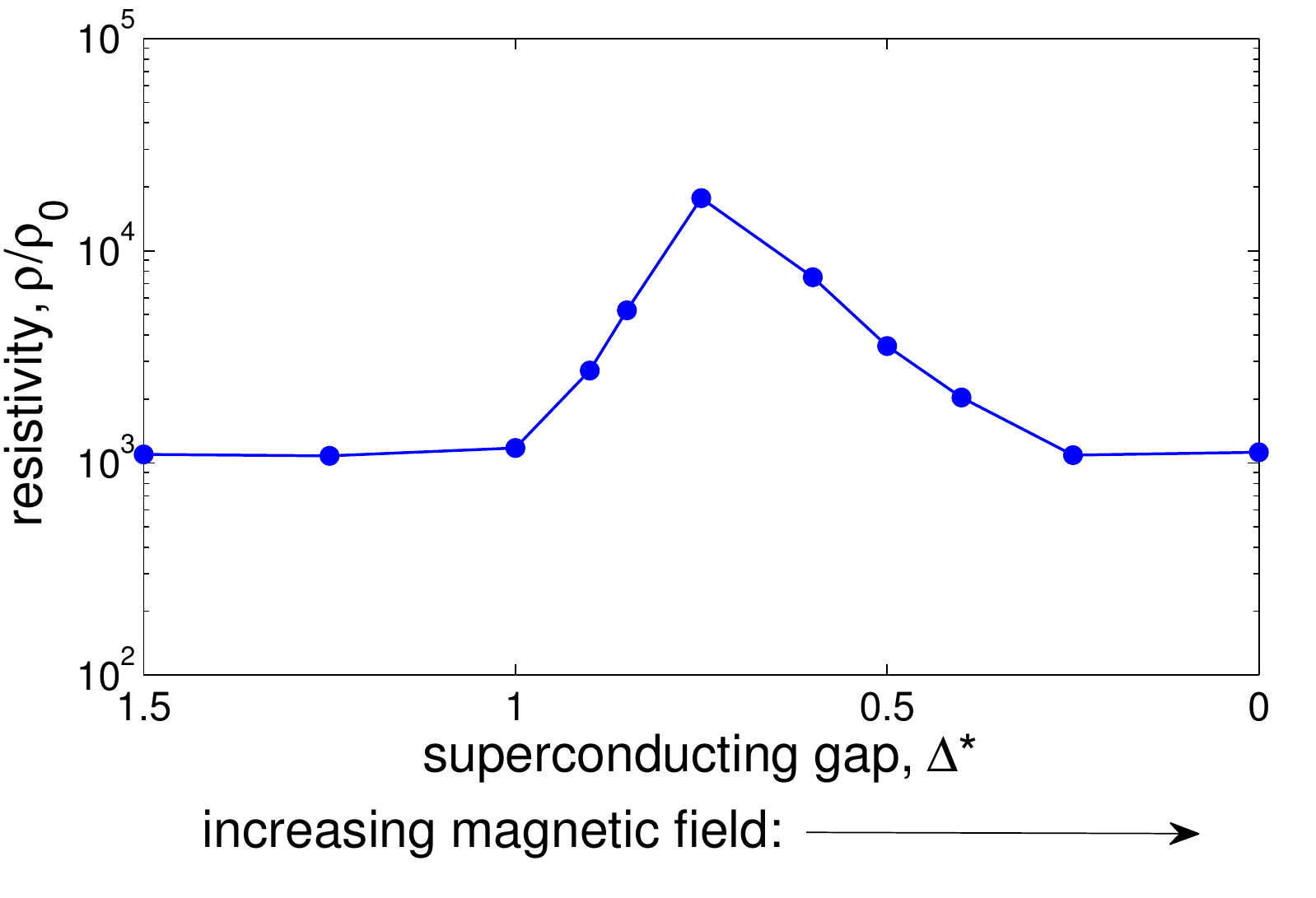}
\caption{(Color online) The resistivity, $\rho/\rho_0$, as a function of the superconducting gap $\Delta^*$ for a 2d array with localization lengths $\xi_1 = D$ and $\xi_2 = 10 D$ and temperature $T = 0.1 E_c/k_B$ [so that $(T^*_1)^{-1/2} = 2.1$ and $(T^*_2)^{-1/2} = 6.7$].  The maximum in $\rho/\rho_0$ suggests a magnetoresistance peak associated with the transition from pair-dominated conduction (at large $\Delta^*$, small magnetic field) to single electron-dominated conduction (at small $\Delta^*$, large magnetic field).}
\label{fig:MRpeak}
\end{figure}

To investigate this possibility, we performed simulations to measure the resistivity at different values of the localization lengths $\xi_1$, $\xi_2$ and the temperature $T$, using a resistor network that allows for mixed conductivity of singles and pairs.  We find that if $\xi_1$, $\xi_2$, and $T$ are chosen such that the resistivities are nearly equal at $\Delta^* \gg 1$ and $\Delta^* = 0$, then one can indeed observe a moderate peak in the resistivity in the vicinity of $\Delta^* = 1$.  One such result is shown in Fig.\ \ref{fig:MRpeak}, and is qualitatively similar to a result obtained in Ref.\ \onlinecite{Mitchell2012tcg} for an array in which the quantity $E_c - \Delta$ varies strongly between grains.

While this result is promising, we caution that by itself it does not provide a satisfactory qualitative description of the magnetoresistance peak observed in experiment.  For example, the peak in Fig.\ \ref{fig:MRpeak} arises out of the deeply insulating state, $\rho \sim 10^3 \rho_0$.  Since the constant $\rho_0$ is generally on the order of $h/e^2 \approx 26$ k$\Omega$, this disagrees with experiment \cite{Steiner2005sii, Lin2011mqp, Baturina2008hri}, where the magnetoresistance peak is seen to arise from a state with $\rho \sim h/e^2$.  The appearance of a noticeable peak also apparently requires a large ratio $\xi_2/\xi_1$, which is likely to be possible only very close to the superconductor-insulator transition.  Such large values of $\xi_2$ probably go beyond the limit of applicability of our model.

\section{3d arrays} \label{sec:3d}

Thus far our presentation of results has focused on the case of 2d arrays.  In this section we briefly report on simulations of the DOGS and resistivity in 3d arrays.  Generally speaking, while some details of the shape of the DOGS and the magnitude of the resistivity are modified relative to the 2d case, the triptych structure of the DOGS and the values of the effective charges remain unchanged.  All results in this section correspond to simulated 3d systems of $24 \times 24 \times 24$ lattice sites with open boundary conditions, averaged over $1000$ realizations of the disorder.

\begin{figure}[htb!]
\centering
\includegraphics[width=0.48 \textwidth]{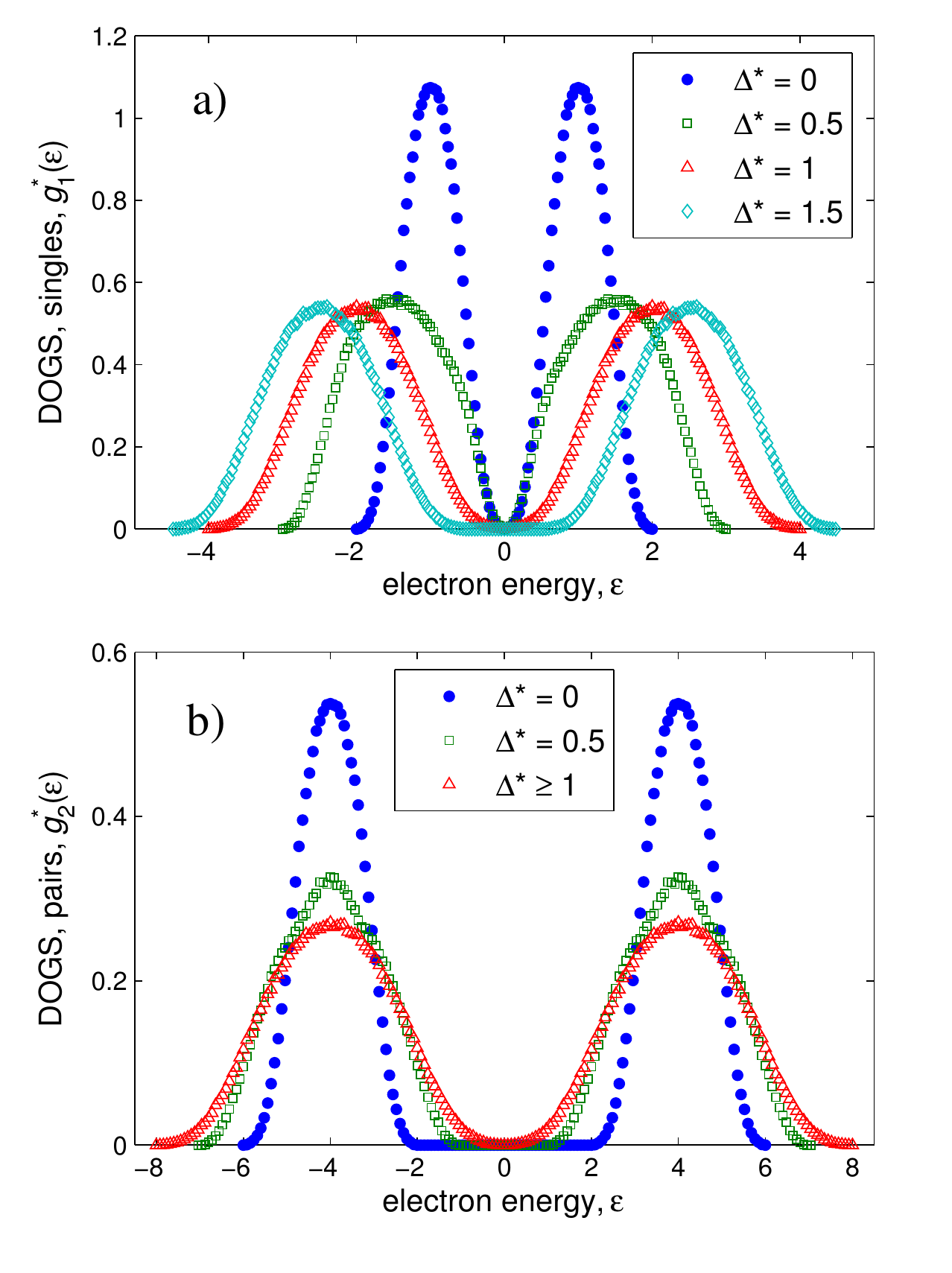}
\caption{(Color online) Single electron and pair DOGS, $g_1^*(\e)$ and $g_2^*(\e)$, of a 3d array of monodisperse metallic grains. The DOGS curves obey the same scaling relations as in 2d, [see Eq.\ (\ref{eq:sqrt2})], indicating the presence of the same effective charges $1e$, $\sqrt{2}e$, and $2e$ and $\Delta^* = 0$, $\Delta^* = 1$, and $\Delta^* > 1$, respectively.}
\label{fig:DOGS3d}
\end{figure}

When considering the DOGS, the most prominent difference between 2d and 3d systems is that in 3d the ES criterion [Eq.\ (\ref{eq:ES1})] imposes a stronger constraint on $g_1^*(\e)$.  Specifically, in $d$ dimensions the ES criterion implies \cite{Efros1984epo} that $g_1^*(\e) < k_d |\e|^{d-1}$, where $k_d$ is a constant, so that in 3d the DOGS vanishes at least quadratically with energy near the Fermi level while in 2d it is constrained to vanish only linearly.  Fig.\ \ref{fig:DOGS3d} shows $g_1^*(\e)$ and $g_2^*(\e)$ for 3d arrays, and one can see that in cases where $g_1^*(\e)$ is ungapped it indeed vanishes as a higher power of $\e$ near the Fermi level.  

Nonetheless, the most important qualitative features of the DOGS from the 2d case remain in 3d as well.  Specifically, the curves corresponding to $g^*_1(\e)$ at $\Delta^* = 0, 1$ and $g^*_2(\e)$ at $\Delta^* \geq 1$ have the ``triptych" structure of two identical, symmetric peaks, and they can be scaled onto each other using the same scaling relations of Eq.\ (\ref{eq:sqrt2}).  This implies that in 3d we have the same effective charges $e^* = 1e$, $e^* = \sqrt{2} e$, and $e^* = 2e$ for single electrons at $\Delta^* = 0, 1$ and for pairs at $\Delta^* \geq 1$, respectively.

Thus, the most important conclusion from our results in 2d remains for the 3d case as well.  This is as expected, since, as explained in Sec.\ \ref{sec:results}, the effective charges arise from the single-electron energy spectrum (Fig.\ \ref{fig:spec}), and are therefore independent of the system dimensionality.

One can also check that in 3d the effective charges have the same influence on the ES temperature as predicted by Eq.\ (\ref{eq:Tesratio}).  By numerically evaluating the resistivity of these 3d systems, we indeed find that the ES temperatures $\Tes$ obey Eq.\ (\ref{eq:Tesratio}).  A plot of the dimensionless resistivity $\ln \rho^*$ against $(T^*)^{-1/2}$ in 3d is essentially identical to that of Fig.\ \ref{fig:Rratio}, with slight downward shifts in the magnitude of the resistivity relative to the 2d case.  Making independent linear fits to the data gives $\Tes^p(\Delta^* \geq 1) \approx 2.6 \Tes^s(\Delta^* = 1) \approx 5.1 \Tes^s(\Delta^* = 0) $, which agrees with Eq.\ (\ref{eq:Tesratio}) to within our numerical uncertainty.

\section{Tunneling experiments} \label{sec:tunneling}

In the previous sections we presented results for the DOGS and we showed that these results have important consequences for the characteristic temperature $\Tes$ and for the magnetoresistance.  In this section we discuss how the DOGS can be observed directly from tunneling experiments.

Tunneling experiments have previously been used to directly observe the Coulomb gap in lightly-doped semiconductors \cite{Lee1999cgi, Massey1995doo}, and have also measured the superconducting gap in isolated superconducting grains \cite{Black1996sos} and in disordered films \cite{Sherman2012mos}.  It is therefore natural to think that the single-electron DOGS $g_1(E)$ predicted here can also be measured via tunneling.  In the problem we are considering, however, the energy scales $E_c$ and $\Delta$ are similar in magnitude, and thus the tunneling conductance reflects a convolution of the DOGS $g_1(E)$ with the density of states $f(E)$ within each grain.  As a result, we consider it worthwhile to explicitly state our predictions for the tunneling conductance $G(\Delta, V)$, where $V$ is the applied voltage, at different values of the gap $\Delta$.  

For simplicity, in this section we ignore the potential effects of spin polarization on the tunneling rates.  This is equivalent to assuming that any applied magnetic field modifies the superconducting gap $\Delta$ primarily through orbital effects rather than the Zeeman effect, so that the electron energy levels shown in Fig.\ \ref{fig:spec} are not labeled by spin.

Since the spacing $\delta$ between discrete electron energy levels within the grain satisfies $\delta \ll E_c$, as explained in the Introduction, we can take the density of states $f(E)$ within each grain to be a continuous function.  For metallic grains with $\Delta = 0$, $f(E)$ can be considered a constant, $f(E) = f_0$, as long as $|(E - \mu)/\mu | \ll 1$.  On the other hand, when $\Delta$ is finite, coherence peaks arise in the density of states \cite{Tinkham1996its}, so that at $|E| > \Delta$
\be 
\frac{f(E)}{f_0} = \frac{E}{\sqrt{ E^2 - \Delta^2 }},
\label{eq:horns}
\ee
where in this expression $E$ is measured relative to the center of the superconducting gap.  [Eq.\ (\ref{eq:horns}) ignores the potential effect of thermal broadening of the coherence peaks.]

The expression of Eq.\ (\ref{eq:horns}) indicates that the conductance into a single grain is greatly enhanced when the voltage is aligned with the edge of the superconducting gap.  For an array of grains, the total conductance is the integrated conductance of all the individual grains, each of which has a different relative alignment with the voltage.  Thus, the differential conductance satisfies 
\be 
G(\Delta, V) = G_0 A D^{1+d} \int_0^{eV} g_1(E) f(eV - E + \Delta) dE,
\label{eq:G}
\ee
where $G_0$ is a constant and $A$ is the area of the tunnel barrier.  [The term $+ \Delta$ in the argument of $f$ in Eq.\ (\ref{eq:G}) accounts for the fact that the function $f(E)$ in Eq.\ (\ref{eq:horns}) is defined relative to the center of the superconducting gap while the ground state energies described by $g_1(E)$ include the gap energy $\Delta$.]  

Given our results for $g_1(E)$, one can use Eq.\ (\ref{eq:G}) to numerically evaluate the conductance $G(\Delta, V)$.  For the limiting case $\Delta = 0$, where the density of states $f(E)$ is constant, Eq.\ (\ref{eq:G}) becomes simply $G(\Delta = 0, V) \propto \int_0^{eV} g_1(E) dE$, or in other words $g_1(eV) \propto dG(0,V)/dV$.  For small but finite $\Delta$, on the other hand, such that $0 < \Delta^* < 1$, the conductance $G(\Delta, V)$ is enhanced at small $V$ relative to $G(0, V)$ as a result of the coherence peaks.  At large $\Delta^* > 1$, a gap opens in $g_1(E)$, and $G(\Delta, V)$ remains at zero for $|eV| <  (\Delta^* - 1)E_c$.

\begin{figure}[htb!]
\centering
\includegraphics[width=0.45 \textwidth]{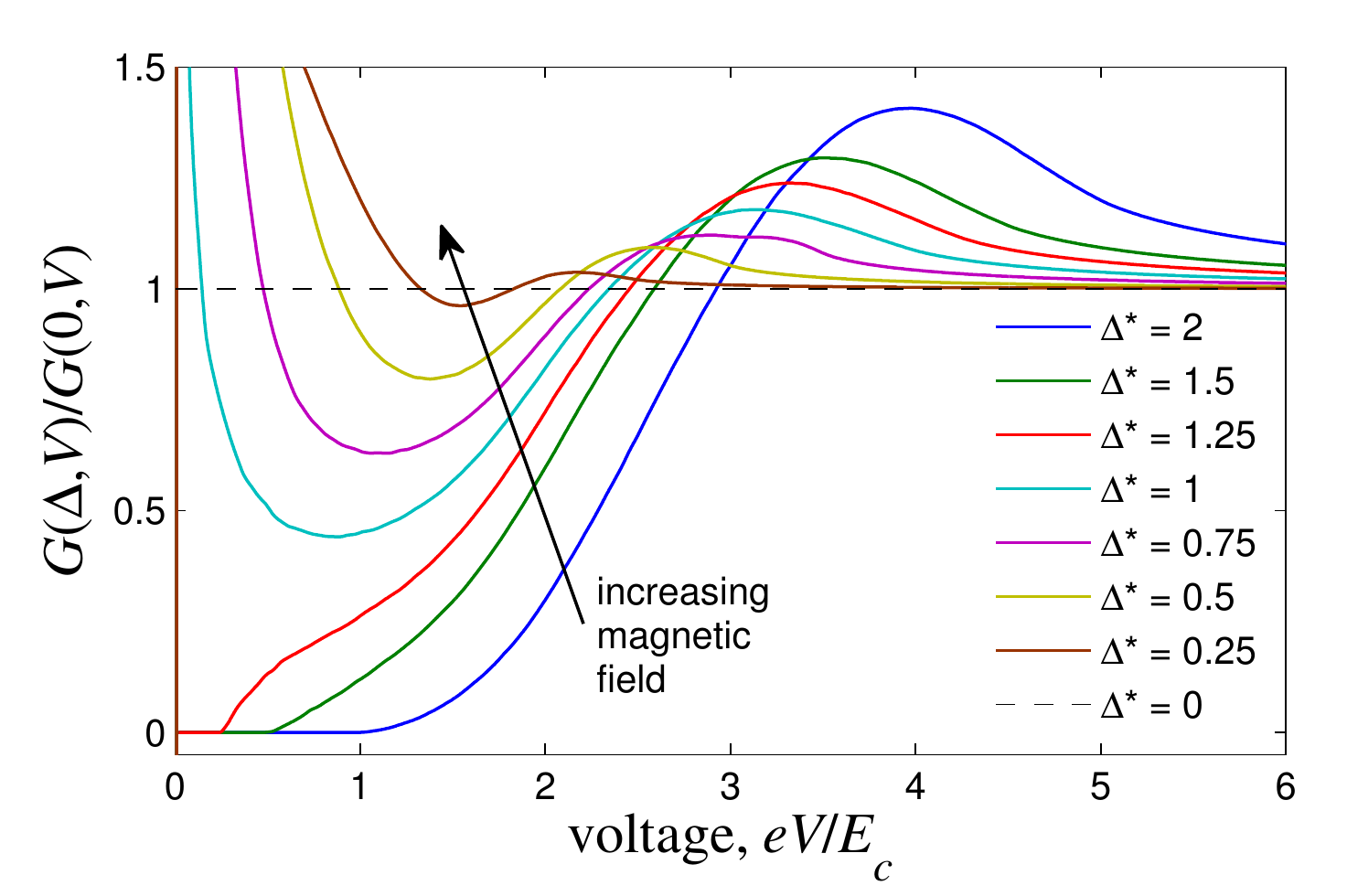}
\caption{(Color online) Tunneling conductance $G(\Delta, V)$ into a 2d array of superconducting grains as a function of voltage $V$ and for different values of the superconducting gap $\Delta^*$, plotted as a ratio of the conductance at $\Delta = 0$.  The voltage $V$ is plotted in the dimensionless form $eV/E_c$.  As the magnetic field is increased, driving down the value of $\Delta$, the ratio $G(\Delta, V)/G(0, V)$ at small voltage changes from zero to a divergently large value as $\Delta^*$ is made smaller than 1.}
\label{fig:G}
\end{figure}

This result is shown in Fig.\ \ref{fig:G} for the case of tunneling into a 2d array.  Here the conductance $G(\Delta, V)$ is plotted normalized to the value $G(0, V)$ as a function of dimensionless voltage $eV/E_c$ for different values of $\Delta^*$.  One can think that these different curves correspond to different magnetic field, since, as explained above, an increased magnetic field reduces the gap $\Delta$.  Thus, Fig.\ \ref{fig:G} suggests that if one starts with a sample for which $\Delta^* > 1$ and increases the magnetic field, a dramatic change occurs in the quantity $G(\Delta, V)/G(0, V)$.  Namely, $G(\Delta, V)/G(0, V)$ first remains at zero for small $V$, since the single-electron DOGS is gapped.  As the magnetic field is increased, the width of this gap decreases, until the point where $\Delta^* = 1$ and it disappears.  Once $\Delta^* \leq 1$, the value of $G(\Delta, V)/G(0, V)$ undergoes an abrupt change such that it becomes divergently large at small voltage.  This divergence can be seen as the result of the coherence peaks, which greatly increase the tunneling at small voltage relative to the case where there is no Cooper pairing (high magnetic field).  Increasing the magnetic field also has the effect of lowering the conductance peak at larger voltage, $eV/E_c \sim 2 + \Delta^*$.
 
It should be noted that the prediction of Fig.\ \ref{fig:G} is dependent on the existence of \emph{unscreened}, long-range Coulomb interactions between electrons, which create the Coulomb gap in $g_1(E)$.  If such long-range interactions are screened by the presence of a nearby metal electrode, which creates an image charge for each charged grain and truncates the $1/r$ interaction, then the Coulomb gap will not be preserved and the predictions of this paper will be modified.  For 2d arrays, it is therefore likely that macroscopic tunneling experiments will not be effective in identifying a Coulomb gap.  The behavior of Fig.\ \ref{fig:G} may nevertheless still be identified if one uses a scanning tunneling tip to measure the conductance through individual grains and takes an ensemble average (as in, e.g., Ref.\ \onlinecite{Sacepe2011lop}).  Alternatively, one can measure the conductance into a 2d face of a thick, 3d array, as was done in Ref.\ \onlinecite{Massey1995doo}.  While Fig.\ \ref{fig:G} plots the conductance assuming tunneling into a 2d array, if one assumes that $g_1^*(\e)$ is identical to that of a 3d system (Fig.\ \ref{fig:DOGS3d}), the results are qualitatively very similar.

\section{Conclusions} \label{sec:conclusion}

In this paper we have proposed a model of a disordered granular superconductor and evaluated the DOGS and resistivity at different values of the superconducting gap $\Delta$.  Our primary result is the DOGS for single electrons and electron pairs shown in Figs.\ \ref{fig:DOGS} and \ref{fig:DOGS3d}.  We also have considered the implications of the DOGS for the conductivity of the system (Figs.\ \ref{fig:Rratio}), and explained a mechanism for negative magnetoresistance (Fig.\ \ref{fig:singlesR}).  Our predictions for the tunneling conductance are given in Fig.\ \ref{fig:G}.  

Perhaps the most remarkable result is the existence of effective charges $1e$, $2e$, and $\sqrt{2}e$ at $\Delta^* = 0$, $\Delta^* > 1$, and $\Delta^* = 1$, respectively, which was first reported by Ref.\ \onlinecite{Mitchell2012tcg}.  These effective charges codify exact scaling relations between different results for the DOGS [Eq.\ (\ref{eq:sqrt2})] and for the conductivity at low temperature [Eq.\ (\ref{eq:Tesratio})], and can be understood in a fairly intuitive way.  At $\Delta^* = 0$, electrons are unpaired and electronic conduction is performed by single electrons.  At $\Delta^* > 1$, electrons become bound in Cooper pairs and these pairs are the primary players both in the conductivity and in determining the DOGS.  At the point $\Delta^* = 1$, single electrons hop in a disorder potential that is shaped primarily by Cooper pairs, and the single-electron DOGS and conductivity can be described by an effective charge $e^* = \sqrt{2}e$.

It is perhaps worth emphasizing that this effective charge $\sqrt{2}e$ does not represent a real quasiparticle in the traditional sense.  For example, unlike the charges $1e$ and $2e$, the charge $\sqrt{2}e$ is unlikely to appear in the shot noise of the current (or the Fano factor), since the actual hopping is performed by single electrons.  Rather, the appearance of the charge $\sqrt{2}e$ in $g_1(E)$ and $\Tes$ is the result of a degeneracy in the electronic spectrum, which results in electrons being paired in the ground state.  These paired electrons rearrange in the presence of a disorder potential and determine the properties of the ground state, while transport is carried by singles.  It is this combination of intimidation by pairs and conduction by singles that produces the appearance of a $\sqrt{2}$ charge.  

More generally, this view represents something of a novel paradigm in hopping transport.  Namely, that a system can be simultaneously populated by two or more charged species (here, singles and pairs), one of which determines the Coulomb landscape while the other is responsible for transport.  Exploring this kind of physics in other classes of disordered systems remains a promising topic for future study.

\begin{acknowledgments}

The authors would like to thank A. Frydman, V. Galitski, Yu. Galperin, A. Gangopadhyay, A. Kamenev, J. Mitchell, M. Mueller, B. Z. Spivak, and A. Vainshtein for helpful discussions.
This work was supported partially by the MRSEC Program of the National Science Foundation under Award Number DMR-0819885.  T. Chen was partially supported by the FTPI.

\end{acknowledgments}

\bibliography{pair_draft}

\end{document}